\documentclass[10pt,conference]{IEEEtran}
\IEEEoverridecommandlockouts
\usepackage{epsfig}
\usepackage{graphicx}
\usepackage{amsmath}
\usepackage{amssymb}               
\usepackage{amsfonts}               
\usepackage{amsopn}

\newcommand{\hY}{\hat{Y}}
\newcommand{\hy}{\hat{y}}

\newcommand{\tX}{\tilde{X}}
\newcommand{\tY}{\tilde{Y}}
\newcommand{\tZ}{\tilde{Z}}
\newcommand{\styp}{\mathit{T}_{\epsilon}^n}

\def\ma{{\mathcal A}}
\def\mb{{\mathcal B}}
\def\mc{{\mathcal C}}

\def\ml{{\mathcal L}}

\def\mn{{\mathcal N}}

\def\ms{{\mathcal S}}
\def\mt{{\mathcal T}}
\def\mu{{\mathcal U}}
\def\mv{{\mathcal V}}
\def\mw{{\mathcal W}}
\def\mx{{\mathcal X}}
\def\my{{\mathcal Y}}
\def\mz{{\mathcal Z}}

\def\um{\mathbf{m}}

\def\bC{\mathbf{C}}

\def\uu{\mathbf{u}}

\def\ux{\mathbf{x}}
\def\uy{\mathbf{y}}

\def\uhy{\mathbf{\hat{y}}}
\newtheorem{theorem}{Theorem}
\newtheorem{lemma}{Lemma}

\newtheorem{definition}{Definition}

\newtheorem{remark}{Remark}

\begin{document}

\author{Mohammad~Hossein~Yassaee, Mohammad~Reza~Aref\\Information Systems and Security Lab (ISSL)\\EE Department, Sharif University of Technology, Tehran, Iran\\
E-mail: yassaee@ee.sharif.edu, aref@sharif.edu
\thanks{\noindent This work was partially supported by Iranian-NSF under grant No. 84.5193-2006}}
\title{Slepian-Wolf Coding over Cooperative Networks}
\maketitle

\begin{abstract}
We present sufficient conditions for multicasting a set of correlated sources over cooperative networks. We propose  \emph{joint source-Wyner-Ziv encoding}/\emph{sliding-window decoding} scheme, in which each receiver considers an ordered partition of other nodes. Subject to this scheme, we obtain a set of feasibility constraints for each ordered partition. We consolidate the results of different ordered partitions by utilizing a result of geometrical approach to obtain the sufficient conditions. We observe that these sufficient conditions are indeed necessary conditions for \emph{Aref networks}. As a consequence of the main result, we obtain an achievable rate region for networks with multicast demands. Also, we deduce an achievability result for two-way relay networks, in which two nodes want to communicate over a relay network. 
\end{abstract}
\fontsize{10}{11}\selectfont
\section{introduction}
We consider the problem of reliable transmission of discrete memoryless correlated sources (DMCS) over cooperative networks in which each node can simultaneously encode a message, relay the messages of other nodes and decode the messages. The main goal of this paper is to find sufficient conditions to the following problem:\par

\emph{ Given a set of sources $U_{\ma}=\{U_{a_j}:a_j\in\ma\}$ observed at nodes $\ma=\{a_1,\cdots,a_M\}\subseteq\mv$ ($\mv=\{1,\cdots,N\}$ is the set of nodes in the network) respectively and a set of receivers at nodes $\mb=\{b_1,\cdots,b_K\}\subseteq\mv$ which is not necessarily disjoint from $\ma$, what conditions must be satisfied to enable us to reliably multicast $U_{\ma}$ to all nodes in $\mb$?} \par

In addition to this problem, we are interested in the special case of reliable transmission of independent sources (messages) over cooperative networks with multicast demands. In particular, we consider the problem of finding a feasible rate region for two-way relay networks as a special case of cooperative networks with two transmitters and two receivers with multicast demands.\par
 
The problem of Slepian-Wolf coding over multi-user channels has been considered for some special networks. In \cite{tuncel}, Tuncel obtained a necessary and sufficient condition for multicasting a source over a broadcast channel with side information at each receiver. He proposed a joint source-channel coding scheme that achieves \emph{operational separation} between source coding and channel coding. In \cite{babu}, a necessary and sufficient condition for multicasting a set of correlated sources over acyclic Aref networks \cite{aref} has been derived. Also the problem of multicasting of correlated sources over networks was studied in network coding literature \cite{ho,effros}.\par
    
Finding the achievable rate region of multi-relay networks is one of the interesting problems in Shannon theory. Based on Decode and Forward strategy, \cite{kramer} and \cite{xie} proposed achievable rates for Multiple Access Relay Channel and multisource, multirelay and multidestination networks, respectively. Compress and Forward (CF) strategy was generalized to relay networks with one source and one destination by several authors in \cite{kramer2005,yassaee}. Also, Avestimehr, et.al in \cite{avesisit08,avesphd} proposed a quantize-map scheme for Gaussian relay networks with multicast demands which achieves the cut-set bound within a constant number of bits. Their scheme is based on Wyner-Ziv encoding at relays and a distinguishability argument at receivers. \par

In this paper, we propose a \emph{joint Source-Wyner-Ziv encoding/sliding window decoding} scheme for Slepian-Wolf coding over cooperative networks. Our scheme results in the operational separation between source and channel coding. In addition, this scheme does not depend on the graph of networks, so the result can easily be applied to any arbitrary network (In general for multi-user networks which are characterized by a conditional probability distribution, it is not always possible to describe networks with a graph). We show that the sufficient conditions, are also necessary conditions for the Slepian-Wolf coding over arbitrary Aref networks. As an another consequence of the proposed scheme, we obtain an achievable rate region based on CF strategy. Moreover, one can easily check that our achievable rate for relay networks subsumes the achievable rates 

obtained for deterministic and Gaussian relay networks in \cite{avesphd}. Finally, we apply the main result and prove an achievability theorem for the two-way relay network, which is consisted of two transmitters communicating over a relay networks.

\section{Preliminaries and Definitions} 
 We denote discrete random variables with capital letters, e.g., $X$, $Y$, and their realizations with lower case letters
$x$, $y$. A random variable $X$ takes values in a set $\mx$. We use $|\mx|$ to denote the cardinality
of a finite discrete set $\mx$, and $p_X(x)$ to denote the probability density function (p.d.f.) of $X$ on $\mx$. For brevity we may omit the subscript $X$ when it is obvious from
the context. We denote vectors with boldface letters, e.g. $\ux$, $\uy$. In addition, we let $X^i=(X_1,\cdots,X_i)$. We use $\styp(X)$ to denote the set of $\epsilon$-strongly typical sequences of length $n$, w.r.t. density
$p_X(x)$ on $\mx$. Further, we use $\styp(Y|\ux)$ to denote the set of all $n$-sequence $\uy$ such that $(\ux,\uy)$ are jointly typical, w.r.t. $p_{XY}(x,y)$. We denote the vectors in the $j$th block by a subscript $[j]$. For a given set $\ms$, we define $X_{\ms}=\{X_i:i\in\ms\}$ and $R_{\ms}=\sum_{i\in\ms}R_i$. \par

We consider the problem of reliable multicasting of the DMCS $U_{\ma}$ to the subset $\mb$ of nodes, where transmission is over discrete memoryless cooperative network $p(y_1,\cdots,y_{N}|x_1,\cdots,x_{N})$ with input alphabet and output alphabet $\mx_v$ and $\my_v$ at each node $v\in\mv$, respectively. A formal definition of the problem is given below.

\begin{definition}
\label{def:sw}
We say that the set of DMCS, $U_\ma$ can reliably be transmitted over discrete memoryless cooperative network to all nodes in $\mb$, if there exist positive integers $(m,n)$ and a sequence of encoding functions 
\[ f_{v,t}^{(m)}:\mu_v^m\times\my_v^{t-1}\rightarrow\mx_v\quad \mbox{for}\quad t=1,\cdots,n\] 
at all nodes $v\in\mv$, where for non-source nodes we let $\mu_v=\emptyset$ 
and a set of decoding functions defined at each node $b_i$;
\[g_{b_i}^{(m,n)}:\mu_{b_i}^m\times\my_{b_i}^n\rightarrow\mu_{\ma}^m\]
such that the probability $Pr(g_{b_i}^{(m,n)}(U_{b_i}^m,Y_{b_i}^n)\neq U_{\ma})$ vanishes for all $b_i\in\mb$ as $m,n$ go to infinity with $\frac{m}{n}$ goes to one.
 
\end{definition}

\section{Summary of Main Results}
In this section, we provide a summary of our main results. The following theorem is the main result of the paper. 
\begin{theorem}
\label{thm:sw}
The set of DMCS $U_{\ma}$ can reliably be transmitted over cooperative network, if there exist auxiliary random variables $\hY_{\mv}$ such that for each $\ms\subseteq\ma$, we have 

\begin{small}
\begin{multline}
\label{eq:sw}
H(U_{\ms}|U_{\ma\backslash\ms})< \min_{b_i\in\mb\backslash\ms}\min_{\mv\supseteq\mw\supseteq\ms: \atop b_i\in\mw^C} [I(X_{\mw};Y_{b_i}\hY_{\mw^C\backslash\{b_i\}}|X_{\mw^C})\\-I(Y_{\mw};\hY_{\mw}|X_{\mv}Y_{b_i}\hY_{\mw^C\backslash\{b_i\}})]
 \end{multline}
 \end{small} 
 \vspace{-1mm}
 \noindent where the joint p.d.f. of random variables 
  factors as
 \begin{equation}
 \label{eq:dist}
 p(u_{\ma})[\prod_{v\in\mv}p(x_v)p(\hy_v|x_v,y_v)]p(y_{\mv}|x_{\mv}).
 \end{equation}
 \end{theorem}
 
 \begin{proof}
 We sketch the proof in the next section.
 \end{proof}
 \begin{remark}
 The constraint \eqref{eq:sw} separates source coding from channel coding in the operational separation sense \cite{tuncel}. The LHS of \eqref{eq:sw} represents the rate of Slepian-Wolf coding, while the RHS of \eqref{eq:sw} provides an achievable flow through a cut $\Lambda=(\mw,\mw^C)$ over the cooperative network. 
 \end{remark} 
 
\par In the rest of this section, we consider some consequences of Theorem \ref{thm:sw}.
First, assume that each channel output is a deterministic function of all channel inputs, i.e., $y_v=g_v(x_{\mv})$.
Setting $\hY_v=Y_v$ in Theorem \ref{thm:sw}, we conclude that the reliable transmission of DMCS over deterministic network is feasible if there exists a product distribution $\prod_v p(x_v)$ such that:
\begin{align}
\label{eq:det}
  H(U_{\ms}|U_{\ma\backslash\ms})< \min_{b_i\in\mb\backslash\ms}\min_{\mv\supseteq\mw\supseteq\ms:\atop b_i\in\mw^C} H(Y_{\mw^C}|X_{\mw^C})
 \end{align}
  In the following lemma, we provide a converse for reliable transmission of correlated sources over deterministic cooperative network.

 \begin{lemma}
 \label{le:ob}
 If a set of DMCS $U_{\ma}$ can reliably be multicast over a deterministic network, then there exists a joint p.d.f. $p(x_{\mv})$ such that 
  
  \begin{equation}
  \label{eq:sw2}
 H(U_{\ms}|U_{\ma\backslash\ms})< \min_{b_i\in\mb\backslash\ms}\min_{\mv\supseteq\mw\supseteq\ms:\atop b_i\in\mw^C} H(Y_{\mw^C}|X_{\mw^C})
 \end{equation}
 \end{lemma}
 
 \begin{proof}
 By Fano's inequality, we have:
 \vspace{-2mm}
 \begin{equation}
 \label{eq:ob}\forall \ms\subseteq\mv, b_i\in\mb\backslash\ms: \frac{1}{m}H(U_{\ms}^m|U_{\ma\backslash\ms}^mY_{b_i}^n)\leq\epsilon
 \end{equation}
 For each $(\mw,b_i)$ such that $\ms\subseteq\mw\subseteq\mv$ and $b_i\in\mw^C$, we have:
 \vspace{-2mm}
 \begin{small}
 \begin{eqnarray}
 H(U_{\ms}|U_{\ma\backslash\ms})&=&\frac{1}{m}H(U_{\ms}^m|U_{\ma\backslash\ms}^m)\nonumber\\
 &=&\frac{1}{m}(I(U_{\ms}^m;Y_{b_i}^n|U_{\ma\backslash\ms}^m)+H(U_{\ms}^m|U_{\ma\backslash\ms}^mY_{b_i}^n))\nonumber\\
 &\leq&\frac{1}{m}I(U_{\ms}^m;Y_{\mw^C}^n|U_{\ma\backslash\ms}^m)+\epsilon\nonumber\\
 &\stackrel{(a)}{=}&\frac{1}{m}H(Y_{\mw^C}^n|U_{\ma\backslash\ms}^m)+\epsilon\nonumber\\
 &\stackrel{(b)}{=}&\frac{1}{m}\sum_{i=1}^{n}H(Y_{\mw^C,i}|U_{\ma\backslash\ms}^mY_{\mw^C}^{i-1}X_{\mw^C,i})+\epsilon\nonumber\\
 &\leq&\frac{1}{m}\sum_{i=1}^{n}H(Y_{\mw^C,i}|X_{\mw^C,i})+\epsilon\nonumber\\
 &\stackrel{(c)}{=}&\frac{n}{m} H(Y_{\mw^C,Q}|X_{\mw^C,Q},Q)+\epsilon\nonumber\\
 &{\leq}&\frac{n}{m} H(Y_{\mw^C,Q}|X_{\mw^C,Q})+\epsilon\nonumber\\
 &\stackrel{(d)}{\rightarrow}&H(Y_{\mw^C}|X_{\mw^C})
 \end{eqnarray}
 
 \end{small}
 
 \noindent where (a) follows because $Y_{\mw^C}^n$ is a function of $U_{\ma}^m$, (b) follows from definition \ref{def:sw}, (c) is obtained by introducing a standard time-sharing random variable $Q$ and (d) follows, by allowing $m,n\rightarrow\infty$ and setting $Y_{\mv}=Y_{\mv,Q}$ and $X_{\mv}=X_{\mv,Q}$.
 
 \end{proof}

 Now we consider two special cases of a deterministic network, linear deterministic finite-field network and Aref network. For linear deterministic finite-field network, it is shown in \cite{avesphd} that the product uniform distribution achieves simultaneously the maximum of RHS of \eqref{eq:sw2} for all $\mw\subseteq\mv$. In Aref network, it is shown that the RHS of \eqref{eq:sw2} only depends on the marginal distributions, i.e., $p(x_v)$. Hence, lemma \ref{le:ob} and \eqref{eq:det} together imply the following theorem:
 \begin{theorem}
  A set of correlated sources can reliably be multicast over a deterministic network, if for each $\ms\subseteq\ma$ the constraint \eqref{eq:det} is satisfied. Moreover, this constraint is indeed necessary for two classes of deterministic networks, namely linear deterministic finite-field network and Aref network.  
 \end{theorem}
 
 \par Now, we concentrate on finding an achievable rate region for cooperative networks. Let $R_v$ be the rate of message of the node $v$. The next theorem gives an achievable rate region for cooperative network.
 \begin{theorem}
 \label{thm:ach}
 An N-tuple $(R_1,R_2,\cdots,R_N)$ is contained in the achievable rate region of cooperative network with multicast demands at each node $b_i\in\mb$, if for each $\ms\subseteq\mv$ the following constraint holds:
  \begin{multline}
\label{eq:ach}
R_{\ms}< \min_{b_i\in\mb\backslash\ms}\min_{\mv\supseteq\mw\supseteq\ms: \atop b_i\in\mw^C} \big[I(X_{\mw};Y_{b_i}\hY_{\mw^C\backslash\{b_i\}}|X_{\mw^C})-\\I(Y_{\mw};\hY_{\mw}|X_{\mv}Y_{b_i}\hY_{\mw^C\backslash\{b_i\}})\big]^+
 \end{multline}
where $[x]^+=\max\{x,0\}$ and the joint p.d.f. of $(x_{\mv},y_{\mv},\hy_{\mv})$ factors as $\prod_{v\in\mv}p(x_v)p(\hy_v|x_v,y_v)]p(y_{\mv}|x_{\mv})$.
 \end{theorem}
 \begin{proof}
  Let $\mt$ be the largest subset of $\mv$ such that the RHS of \eqref{eq:sw} is nonnegative  subject to each $\ms\subseteq\mt$ (Note that if two subsets $\mt_1,\mt_2$ have this property, then $\mt_1\cup\mt_2$ also has this property, so such $\mt$ is unique). Now let $\ma=\mt$ in Theorem \ref{eq:sw}. Assume $U_v\ (v\in\ma)$ have uniform distribution over the set $\mu_v$ and be mutually independent. Substituting $R_v=H(U_v)$ in Theorem \ref{thm:sw} yields that $U_{\mt}$ can reliably be multicast, if \eqref{eq:ach} holds. Hence $(R_1,\cdots,R_N)$ is achievable (Note that $R_v=0$ for each node $v\in\mt^C$).
 \end{proof} 

\begin{remark}
\label{rem:1}
Consider a relay network with node $1$ as a transmitter which has no channel output, i.e., $Y_1=\emptyset$, $N-2$ relay nodes $\{2,\cdots,N-1\}$ and node $N$ as a destination which has no channel input, i.e., $X_N=\emptyset$. Substituting $R_2=\cdots=R_{N}=0$ in Theorem \ref{thm:ach} gives the following achievable rate ($R_{CF}$) for relay network.  
\begin{multline}
\label{eq:ach:rel}
R_{CF}=\min_{\ms\subseteq\mv:\atop 1\in\ms,N\in\ms^C}\big[I(X_{\ms};\hY_{\ms^C\backslash\{N\}}Y_N|X_{\ms^C})-\\I(Y_{\ms};\hY_{\ms}|X_{\mv}Y_N\hY_{\ms^C\backslash\{N\}})\big]^+
\end{multline}
It can be shown that this rate subsumes the achievable rate of \cite[Theorem 3]{yassaee}.
\end{remark}
\begin{remark}
Consider a two-way relay network with nodes $1$ and $N$ as two transmitters, each demanding the message of the other node, and $N-2$ relay nodes $\{2,\cdots,N-1\}$. Substituting $R_2=\cdots=R_{N-1}=0$ and $\hY_1=\hY_N=\emptyset$ in Theorem \ref{thm:ach} gives the following achievable rate region for two-way relay network.
\vspace{-3mm}  
\begin{multline}
k=1,N:\  R_{k}=\min_{\ms\subseteq\mv:\atop k\in\ms,\bar{k}\in\ms^C}\big[I(X_{\ms};\hY_{\ms^C\backslash\{\bar{k}\}}Y_{\bar{k}}|X_{\ms^C})-\\I(Y_{\ms\backslash\{k\}};\hY_{\ms\backslash\{k\}}|X_{\mv}Y_{\bar{k}}\hY_{\ms^C\backslash\{\bar{k}\}})\big]^+
\vspace{-2mm}
\end{multline}

\noindent where $\bar{1}=N$ and $\bar{N}=1$.
\end{remark}
\begin{remark}
Suppose the channel output of relay nodes be a function of channel inputs, i.e., $\forall v\in\mv\backslash\{1,N\}:\ y_v=g_v(x_{\mv})$. Set $\hy_v=y_v$ in \eqref{eq:ach:rel}, we deduce that the cut-set bound is achievable for product distribution. This is a generalization of \cite[Theorem 4.2.3]{avesphd} which states that cut-set bound is achievable under product distribution for deterministic network.
\end{remark}
\begin{remark}
In \cite{avesisit08,avesphd}, authors show that by quantization at noise level, Gaussian relay network achieves the cut-set bound within $5|\mv|$ bits. It can be shown using \cite[Appendix A.5]{avesphd} and quantization at the noise level that the achievable rate of Remark \ref{rem:1} achieves the cut-set bound within $\left\lfloor\frac{3}{2} |\mv|\right\rfloor-1$ bits. A similar result holds for two-way Gaussian relay network.
\end{remark}

\section{proof of Theorem 1}
We prove Theorem \ref{thm:sw} in three steps. In subsection \ref{sub:a}, we propose a joint source-Wyner-Ziv encoding/sliding window decoding scheme. For encoding, each node first compresses its observation using Wyner-Ziv coding, then jointly maps its source sequence and compressed observation to a codeword. In the decoding part of the scheme, each receiver considers an ordered partition of other nodes to decode jointly the sources and the compressed observations of other nodes. We provide a set of sufficient conditions for reliable transmission of DMCS over cooperative networks. In subsection \ref{sub:b}, by applying a result of geometrical approach \cite{yassaee}, we unify the results of subsection \ref{sub:a} under different ordered partitions. The result of this section, yields Theorem \ref{thm:sw} with an additional set of constraints corresponding to reliable decoding of the compressed observations of other nodes. In subsection \ref{sub:c}, we show that without loss of generality, we can neglect these constraints. This completes the proof.
\vspace{-1mm}
\subsection{Joint Source-Wyner-Ziv coding/Sliding Window Decoding}
We transmit $m=nB$ length source over cooperative network in $B+2V-3$ blocks of length $n$ where $V$ is the cardinality of $\mv$.

\subsubsection*{Codebook generation at node $v$} Fix $\delta>0$ such that $|\styp(U_v)|<2^{n(H(U_{v})+\delta)}$. To each element of $\styp(U_v)$, assign a number $w_v\in[2,2^{n(H(U_{v})+\delta)}]$ using a one-to-one mapping. Moreover, we assign one to each non-typical sequence $\uu_v$. We denote the result by $\uu_{v}(w_v)$. For channel coding repeat independently the following procedure $V$ times. We denote the resulting $k$th codebook by $\mc_v(k)$.\\ Choose $2^{n(H(U_{v})+I(Y_v;\hY_v|X_v)+2\delta)}$ codewords $\ux_v(w_v,z_v)$, each drawn uniformly and independently from the set $\styp(X_v)$ where $z_v\in[1,2^{n(I(Y_v;\hY_v|X_v)+\delta)}]$. For Wyner-Ziv coding, for each $\ux_v(w_v,z_v)$ create $2^{n(H(U_v)+\delta)}$ lists $\mathbf{L}_{v}(w'_v)$ with $2^{n(I(Y_v;\hY_v|X_v)+\delta)}$ codewords each drawn uniformly and independently from the set $\styp(\hY_v|\ux_v)$ where $w'_v\in[1,2^{n(H(U_{v})+\delta)}]$. We denote the codewords of $\mathbf{L}_v(w'_v)$ by $\uhy_v(w'_v,z'_v|\ux_v)$ where $z'_v\in[1,2^{n(I(Y_v;\hY_v|X_v)+\delta)}]$.   
\label{sub:a}
\subsubsection*{Encoding at node $v$}
Divide the $nB$-length source stream $u_v^{nB}$ into $B$ vectors $(\uu_{v,[j]}:1\leq j\leq B)$ where $\uu_{v,[j]}=(u_{v,(j-1)n+1},\cdots,u_{v,jn})$. We say that channel encoder receives $\um_v=(m_{v,[1]},\cdots,m_{v,[B]})$, if for $1\leq j\leq B$, $\uu_{v,[j]}$ was assigned to $m_{v,[j]}\in[1,2^{n(H(U_{v})+\delta)}]$. Encoding performs in $B+2V-3$ blocks where in block $b$, we use the codebook $\mc_v(b\mod V)$. For $1\leq b\leq B+2V-3$, define:
\[
 w_{v,[b]}= \left\{
            \begin{array}{ll}
                m_{v,[b-V+1]} &, V\le b\le B+V-1\\
                1 & ,\mbox{otherwise}
            \end{array}\right.
            \]
 In block $1$, a default codeword, $\ux_v(1,1)$ is transmitted. In block $b>1$, by knowing $z_{v,[b-1]}$ from Wyner-Ziv coding at the end of block $b-1$ (described below), node $v$ transmits $\ux_v(w_{v,[b]},z_{v,[b-1]})$.

\begin{table*}

\centering
    \caption{Encoding Scheme for Multicast over network with $\mv=\{1,2,3,4\}$ (The encoding scheme of other nodes is similar)}
    \label{ta:enc}
    \vspace{-0.6cm}
    
       \begin{tabular}[t]{|c|c|c|c|c|c|c|c|}
               \hline
               Node &Block 1& Block 2& Block 3& Block 4& Block 5& Block 6& Block 7 \\
                \hline \hline
                &  &  &  & $\uu_1(m_{1[1]})$ & $\uu_1(m_{1[2]})$ &  &  \\
                 1 & $\ux_1(1,1)$ & $\ux_1(1,z_{1[1]})$ & $\ux_1(1,z_{1[2]})$ & $\ux_1(m_{1[1]},z_{1[3]})$ & $\ux_1(m_{1[2]},z_{1[4]})$ & $\ux_1(1,z_{1[5]})$ & $\ux_1(1,z_{1[6]})$ \\
                & $\uhy_1(1,z_{1[1]}|\ux_{1[1]})$ & $\uhy_1(1,z_{1[2]}|\ux_{1[2]})$ & $\uhy_1(m_{1[1]},z_{1[3]}|\ux_{1[3]})$ & $\uhy_1(m_{1[2]},z_{1[4]}|\ux_{1[4]})$ & $\uhy_1(1,z_{1[5]}|\ux_{1[5]})$ & $\uhy_1(1,z_{1[6]}|\ux_{1[6]})$ & $\uhy_1(1,z_{1[7]}|\ux_{1[7]})$\\
                \hline
                 &  &  &  & $\uu_2(m_{2[1]})$ & $\uu_2(m_{2[2]})$ &  &  \\
                 2 & $\ux_2(1,1)$ & $\ux_2(1,z_{2[1]})$ & $\ux_2(1,z_{2[2]})$ & $\ux_2(m_{2[1]},z_{2[3]})$ & $\ux_2(m_{2[2]},z_{2[4]})$ & $\ux_2(1,z_{2[5]})$ & $\ux_2(1,z_{2[6]})$ \\
                & $\uhy_2(1,z_{2[1]}|\ux_{2[1]})$ & $\uhy_2(1,z_{2[2]}|\ux_{2[2]})$ & $\uhy_2(m_{2[1]},z_{2[3]}|\ux_{2[3]})$ & $\uhy_2(m_{2[2]},z_{2[4]}|\ux_{2[4]})$ & $\uhy_2(1,z_{2[5]}|\ux_{2[5]})$ & $\uhy_2(1,z_{2[6]}|\ux_{2[6]})$ & $\uhy_2(1,z_{2[7]}|\ux_{2[7]})$\\
                \hline
       \end{tabular}

\end{table*}

 \subsubsection*{Wyner-Ziv coding} 
 At the end of block $b$, node $v$ knows $(\ux_{v,[b-1]},\uy_{v,[b-1]})$ and $w_{v,[b]}$ (note that $\um_v$ is available non-causally at node $v$), considers the list $\mathbf{L}_v(w_{v,[b]})$ and declares that $z_{v,[b-1]}=z_v$ is received if $z_{v}$ is the smallest index such that $(\uhy_{v,[b-1]}(w_{v,[b]},z_{v}|\ux_{v,[b-1]}),\ux_{v,[b-1]},\uy_{v,[b-1]})$ are typical. Since $\mathbf{L}_v(w_{v,[b]})$ contains $2^{n(I(Y_v;\hY_v|X_v)+\delta)}$ codewords, such $z_v$ exists with high probability (See Table \ref{ta:enc} which describes encoding for network with four nodes).
\subsubsection*{Decoding at node $b_i$}
Let $\bC^{(b_i)}=[\ml_1,\cdots,\ml_{\ell}]$ be an ordered partition of the set $\mv_{-b_i}=\mv\backslash\{b_i\}$. We propose a sliding window decoding with respect to $\bC^{(b_i)}$. Define $s_{v,[b]}=(w_{v,[b]},z_{v,[b-1]})$. Suppose that $(s_{\ml_1,[b-1]},s_{\ml_2,[b-2]},\cdots,s_{\ml_\ell,[b-\ell]})$ were decoded correctly at the end of block $b-1$. The node $b_i$, declares that $(s_{\ml_1,[b]},\cdots,s_{\ml_{\ell},[b-\ell+1]})=(\hat{s}_{\ml_1},\cdots,\hat{s}_{\ml_{\ell}})$ was sent, if for each $1\le k\le\ell+1$,
\begin{small}
\begin{multline}
\label{eq:typ}
 \mbox{if}\ 1\le b-k+1:\Big(\ux_{\ml_k}(\hat{s}_{\ml_k}),\uhy_{\ml_{k-1}}(\hat{s}_{\ml_{k-1}}|\ux_{\ml_{k-1},[b-k+1]})\\,
\ux_{\ml^k,[b-k+1]},\uhy_{\ml^{k-1},[b-k+1]},\uy_{b_i,[b-k+1]},\ux_{b_i,[b-k+1]}\Big)\in\styp\\
\noindent {\mbox{if $V\le b-k+1\le V+B$:}\  }(\uu_{\ml_k}(\hat{w}_{\ml_k}),\\ \uu_{\ml^k}(w_{\ml^k,[b-k+1]}), \uu_{b_i,[b-k+1]})\in\styp
\end{multline}
\end{small}
where $\ml^k=\cup_{j=1}^{k-1}\ml_j$, $\ml_0=\ml_{\ell+1}=\emptyset$ and $s_{\ml_k}=(w_{\ml_k},z_{\ml_k})$.\\
Note that at the end of block $B+V+\ell-2$, the vector $\um_{\ma}$ is decoded. Since each $(\uu_{v,[j]}:v\in\ma,1\le j\le B)$ is typical with high probability, we find the source sequence $u_{\ma}^{nB}$ with small probability of error. 
\subsubsection*{ Error Probability Analysis}
We bound the probability of error in \eqref{eq:typ} as follows:
\begin{multline}
\label{eq:sum}
\mathbb{P}_e=\sum_{\emptyset\neq\ms\subseteq\mv_{-b_i}}\sum_{\mw\subseteq\ms}P \Big(\exists(\hat{s}_{\ml_1},\cdots,\hat{s}_{\ml_{\ell}})\in\\ \mn_{\ms\mw}^{(1)}\times\cdots\times\mn_{\ms\mw}^{(\ell)}: (\hat{s}_{\ml_1},\cdots,\hat{s}_{\ml_{\ell}})\ \mbox{satisfies \eqref{eq:typ}}\Big)
\end{multline}
where $\mn_{\ms,\mw}^{(k)}$ is the following set:
\begin{small}
\begin{multline*}
 \mn_{\ms,\mw}^{(k)}=\{s_{\ml_k}: \forall t\in\ms_k \ \mbox{and}\ t'\in\mw_k ,s_{t}\neq s_{t,[b-k+1]}\\w_{t'}\neq w_{t',[b-k+1]}\ \mbox{and} \  s_{\ms_k^C}=s_{\ms^C_k,[b-k+1]}, w_{\mw_k^C}=w_{\mw_k^C,[b-k+1]}\}
\end{multline*}
\end{small} 
where  $\ms_k=\ms\cap\ml_k$, $\mw_k=\mw\cap\ml_k$, $\ms^C_k=\ms^C\cap\ml_k$ and $\mw^C_k=\mw^C\cap\ml_k$. \par The probability inside the summation \eqref{eq:sum} represents the probability of error corresponding to incorrect decoding of $s_{\ms}$ such that $w_{\ms\backslash\mw}$ was decoded correctly. Denote this probability by $\mathbb{P}_{e,\ms,\mw}$. We compute it in equation \eqref{eq:prod} shown at the top of the next page, in which (a) follows, because $\ell\le V$ and the codebook generation of any $V$ consecutive blocks are independent. Moreover, the codebook generation is independent of source stream and the sources are i.i.d., so the source sequences are generated independently in consecutive blocks. (b) follows from the fact that $\ux_{t}(s_t)$ and $\uhy_{t}(s_t|\ux_t)$ were drawn uniformly and independently from the sets $\styp({X_t})$ and $\styp(\hY_t|\ux_t)$, respectively.

\begin{figure*}

\begin{align}
\label{eq:prod}
\mathbb{P}_{e,\ms,\mw}\stackrel{(a)}{=}&\sum_{(\hat{s}_{\ml_1},\cdots,\hat{s}_{\ml_{\ell}})\atop\in \mn_{\ms\mw}^{(1)}\times\cdots\times\mn_{\ms\mw}^{(\ell)}}\prod_{k=1}^{\ell+1}\Big[P\big((\ux_{\ml_k}(\hat{s}_{\ml_k}), \uhy_{\ml_{k-1}}(\hat{s}_{\ml_{k-1}}|\ux_{\ml_{k-1},[b-k+1]}), \nonumber
\ux_{\ml^k,[b-k+1]},\uhy_{\ml^{k-1},[b-k+1]},\uy_{b_i,[b-k+1]},\ux_{b_i,[b-k+1]})\in\styp\big)\\& \qquad \qquad \quad \quad \quad \quad \times P\big((\uu_{\ml_k}(\hat{w}_{\ml_k}), \uu_{\ml^k}(w_{\ml^k,[b-k+1]}),\uu_{b_i,[b-k+1]})\in\styp\big)\Big]\nonumber\\
\stackrel{(b)}{=}&\prod_{p=1}^{\ell}|\mn_{\ms\mw}^{(p)}|\prod_{k=1}^{\ell+1}
\big(\frac{|\styp(X_{\ms_k},\hY_{\ms_{k-1}}|\ux_{\ms^C_k}, \uhy_{\ms^C_{k-1}},\ux_{\ml^k},\uhy_{\ml^{k-1}},\uy_{b_i},\ux_{b_i})|}{\prod_{t\in\ms_k}|\styp(X_t)|\prod_{t'\in\ms_{k-1}}|\styp(\hY_{t'}|\ux_{t'})|} \times
\frac{|\styp(U_{\mw_k}|\uu_{\mw_k^C}\uu_{\ml^k}\uu_{b_i})|}{\prod_{t\in\mw_k}|\styp(U_t)|}\big)
\end{align}
\hrulefill
\end{figure*}

 Note that each $(z_t:t\in\ \ms)$ and $(w_{t'}:t'\in\mw)$ take $2^{n(I(Y_t;\hY_t|X_t)+\delta)}$ and $2^{n(H(U_{t'})+\delta)}$ values, respectively. This fact, \eqref{eq:sum} and \eqref{eq:prod} together imply that for reliable decoding, for each $(\ms,\mw)$ such that $\mw\subseteq\ms$, we must have:
\vspace{-.2cm}
\begin{small}
\begin{multline}
\label{eq:suff}
\sum_{t\in\ms}I(Y_t;\hY_t|X_t)\leq\sum_{t\in\ms}H(X_t\hY_t)-
\sum_{k=1}^{\ell+1}\big(H(U_{\mw_k}|U_{\mw_k^C}U_{\ml^k}U_{b_i})\\+H(X_{\ms_k}\hY_{\ms_{k-1}}|X_{\ms_k^C}\hY_{\ms_{k-1}^C}X_{\ml^k}\hY_{\ml^{k-1}}Y_{b_i}X_{b_i})\big)
 \end{multline}   
\end{small}
Note that the RHS of \eqref{eq:suff} takes the minimum value for $\mw=\ms$. Hence we proved the following lemma:
\begin{lemma}
\label{le:first}
The set of DMCS $U_{\ma}$ can reliably be multicast over cooperative network to the subset $\mb$ of nodes, if for each $b_i\in\mb$, there is an ordered partition $\mathbf{C}^{(b_i)}$ of $\mv\backslash\{b_i\}$ such that for each $\ms\subseteq\mv_{-b_i}$, the following constraint holds:   
\vspace{-.3cm}
\begin{small}
\begin{multline}
\label{eq:sufficient}
\sum_{t\in\ms}H(X_t\hY_t)-I(Y_t;\hY_t|X_t)\geq
\sum_{k=1}^{\ell+1}\big(H(U_{\ms_k}|U_{\ms_k^C}U_{\ml^k}U_{b_i})\\+H(X_{\ms_k}\hY_{\ms_{k-1}}|X_{\ms_k^C}\hY_{\ms_{k-1}^C}X_{\ml^k}\hY_{\ml^{k-1}}Y_{b_i}X_{b_i})\big)
 \end{multline} 
 \end{small}
where random variables $(x_{\mv},y_{\mv},\hy_{\mv})$ are distributed according to \eqref{eq:dist}. 
\end{lemma}
\begin{remark}
If there is only one destination, one can use offset encoding scheme \cite{kramer,xie} which has less delay than the proposed encoding scheme, to prove lemma \ref{le:first}. But in general, since the ordered partitions corresponding to each receiver for reliable decoding are different, it is not possible to obtain a same offset encoding scheme for all destinations. This makes clear why the encoding scheme does not transmit any information in the first $V-1$ blocks.  
\end{remark}
\begin{remark}
In the error analysis, we only compute the error corresponding to block $V+\ell-1\le b\le V+B$, for which all $\ell$ consecutive blocks $(b-\ell+1,\cdots,b)$ contain sources' information. However, it can be shown that the constraints are obtained from error analysis of other blocks which correspond to blocks that do not have information about the sources, is dominated by \eqref{eq:sufficient}.
\end{remark}  
\subsection{Unified Sufficient Condition}
\label{sub:b}
In this subsection, we provide a set of sufficient conditions that do not depend on a specified ordered partition. To do this, we need the following lemma which was partially stated in \cite{yassaee} as a result of geometrical properties of achievable rate regions obtained from sequential decoding:
\begin{lemma}
\label{le:geo}
Let $\mathbf{F}_{\mz}$ be the collection of all ordered partitions of a set $\mz$. For each $\mathbf{C}=[\ml_1,\cdots,\ml_{\ell}]\in\mathbf{F}_{\mz}$, define 
\vspace{-.3cm}
\begin{multline}
\label{eq:geo}
\mathbf{R}_{\mathbf{C}}= \{(R_1,\cdots,R_{|\mz|})\in\mathbb{R}^{|\mz|}:\forall\ms\subseteq\mz\\
R_{\ms}\ge\sum_{k=1}^{\ell+1}H(\tY_{\ms_{k-1}}\tX_{\ms_k}|\tX_{\ms^C_k}\tY_{\ms^C_{k-1}}\tX_{\ml^k}\tY_{\ml^{k-1}}\tZ)\}
\end{multline}
then for any joint distribution $p(\tilde{x}_{\mz},\tilde{y}_{\mz},\tilde{z})$, the following identity holds: 
\vspace{-.3cm}
\begin{multline}
\label{eq:geo2}
\bigcup_{\mathbf{C}\in\mathbf{F}_{\mz}}\mathbf{R}_{\mathbf{C}}=\{(R_1,\cdots,R_{|\mz|})\in\mathbb{R}^{|\mz|}:\forall\ms\subseteq\mz\\
R_{\ms}\ge H(\tY_{\ms}\tX_{\ms}|\tX_{\ms^C}\tY_{\ms^C}\tZ)\}  
\end{multline}

\end{lemma}
\begin{proof}
The proof is omitted due to the space limitation. 
\end{proof}
Now consider the RHS of \eqref{eq:sufficient}. Since the random variables $(U_{\ma})$ and $(X_{\mv},\hY_{\mv},Y_{\mv})$ are independent, the RHS of \eqref{eq:sufficient} can be expressed in the form of \eqref{eq:geo} with $\mz=\mv_{-b_i}$, $\tX_t=(X_t,U_t)$, $\tY_t=\hY_t$ and $\tZ=(Y_{b_i},X_{b_i},U_{b_i})$. For each $v\in\mv$, define $R_v=H(X_v\hY_v)-I(Y_v;\hY_v|X_v)$ and let 
\[R^{(b_i)}=(R_1,\cdots,R_{b_i-1},R_{b_i+1},\cdots,R_V)\]
Lemma \ref{le:first} states that $U_{\ma}$ can be multicast over the network, if for each $b_i$ there exists $\mathbf{C}^{(b_i)}\in\mathbf{F}_{\mv_{-b_i}}$ such that $R^{(b_i)}\in\mathbf{R}_{\mathbf{C}^{(b_i)}}$. Applying lemma \ref{le:geo}, we conclude that such $\mathbf{C}^{(b_i)}$ exists iff :
\vspace{-.1cm}
\begin{align}
\forall b_i\in\mb,&\ \ms\subseteq\mv_{-b_i} :\nonumber\\
\label{eq:uni}
R^{(b_i)}_{\ms} \ge& H(\hY_{\ms}X_{\ms}|X_{\ms^C}\hY_{\ms^C}Y_{b_i}X_{b_i})+H(U_{\ms}|U_{\ms^C}U_{b_i}) 
\end{align}
 Note that we can write \eqref{eq:uni} in the following form which will be used in the subsection \ref{sub:c} to complete the proof of Theorem \ref{thm:sw}:
\vspace{-.2cm}
\begin{small}
 \begin{align}
\forall  \ms\subseteq\ma\backslash\{b_i\} :&\nonumber\\
H(U_{\ms}|U_{\ma\backslash\ms})\le & \min_{\mw\supseteq\ms\atop b_i\in\mw^C} R^{(b_i)}_{\mw}-H(\hY_{\mw}X_{\mw}|X_{\mw^C}\hY_{\mw^C\backslash\{b_i\}}Y_{b_i})\nonumber\\ \label{eq:adi}
\forall \ms\subseteq\ma^C\backslash\{b_i\}:
&R^{(b_i)}_{\ms} - H(\hY_{\ms}X_{\ms}|X_{\ms^C}\hY_{\ms^C\backslash\{b_i\}}Y_{b_i})\ge 0 
\end{align}
\vspace{-8mm}   
\end{small}
\subsection{Final Result}
\label{sub:c}
This subsection claims that for each $b_i$, we can reduce the constraints of \eqref{eq:adi} to the first term of it. We prove this by induction on $|\mv_{-b_i}|$. If $|\mv_{-b_i}|=1$, there is nothing to prove. Now suppose the induction assumption is true for all $k<|\mv_{-b_i}|$. 
For each $\mz\subseteq\mv$ which contains $b_i$ and each $\ms\subseteq\mz\backslash\{b_i\}$, let
\vspace{-.1cm} 
\[h^{(b_i)}_{\mz}(\ms)=R^{(b_i)}_{\ms} - H(\hY_{\ms}X_{\ms}|X_{\mz\backslash\ms}\hY_{\mz\backslash(\ms\cup\{b_i\})}Y_{b_i})\] 
Assume there is a subset $\mt$ of $\ma^C\backslash\{b_i\}$ such that $h^{(b_i)}_{\mv}(\mt)<0$. For each $\mw\subseteq\mv_{-b_i}$ observe that,
\begin{small}
\vspace{-.1cm}
\begin{IEEEeqnarray}{rLl}
h^{(b_i)}_{\mv}(\mw\cup\mt)&=&h^{(b_i)}_{\mv}(\mt)+R^{(b_i)}_{\mw\backslash\mt}-\nonumber\\
&&\qquad H(\hY_{\mw}X_{\mw}|X_{\mw^C\backslash\mt}\hY_{\mw^C\backslash(\mt\cup\{b_i\})}Y_{b_i}) \nonumber\\
&<&R^{(b_i)}_{\mw\backslash\mt}-H(\hY_{\mw}X_{\mw}|X_{\mw^C\backslash\mt}\hY_{\mw^C\backslash(\mt\cup\{b_i\})}Y_{b_i}) \nonumber\\
&\le&R^{(b_i)}_{\mw}-H(\hY_{\mw}X_{\mw}|X_{\mw^C}\hY_{\mw^C\backslash\{b_i\}}Y_{b_i}) \nonumber\\  \label{eq:symp}
&=&h^{(b_i)}_{\mv}(\mw)
\end{IEEEeqnarray}
\end{small} 
Using \eqref{eq:symp}, the first term of \eqref{eq:adi} can be simplified as follows:

\begin{IEEEeqnarray}{rCl}
H(U_{\ms}|U_{\ma\backslash\ms})&\le& \min_{\mv\supset\mw\supseteq\ms:\atop b_i\in\mw^C} h^{(b_i)}_{\mv}(\mw)\nonumber\\
&\stackrel{(a)}{=}&\min_{\mv\supset\mw\supseteq\ms:\atop b_i\in\mw^C} h^{(b_i)}_{\mv}(\mw\cup\mt)\nonumber\\
&\stackrel{(b)}{\le}&\min_{\mv\supset\mw\supseteq\ms:\atop b_i\in\mw^C}h^{(b_i)}_{\mv\backslash\mt}(\mw\backslash\mt)\nonumber\\
\label{eq:fin}
&=& \min_{\mv\backslash\mt\supset\mw\supseteq\ms:\atop b_i\in\mw^C}h^{(b_i)}_{\mv\backslash\mt}(\mw)
\end{IEEEeqnarray}  
where (a) follows from \eqref{eq:symp}, because $\ms\subset\mw\cup\mt$ and $b_i\notin\mt$ and (b) follows from the first inequality in \eqref{eq:symp}.\\
Now by induction assumption, the last term of \eqref{eq:fin} corresponds to the feasibility constraints of reliable transmission of $U_{\ma}$ to node $b_i$ over cooperative network with the set of nodes $\mv\backslash\mt$. Hence $U_{\ma}$ can reliably be transmitted to node $b_i$ over original network. This proves our claim. Now it is easy to see that the first term of \eqref{eq:adi} is equivalent to \eqref{eq:sw}, that proves Theorem \ref{thm:sw}.

\section{conclusions}
This paper obtained sufficient conditions for multicasting a set of correlated sources over a cooperative network. The sufficient conditions resulted in an operational separation between source and channel coding. It was shown that these sufficient conditions are also necessary for the Aref network. As a special case, an achievable rate region for cooperative network was derived and the result was specified to the relay network and two-way relay network. Moreover, it was partially shown that these achievable rate regions subsume some recent achievable rate regions which were derived using Wyner-Ziv coding.     
\section{Acknowledgement}
The authors wish to thank M. B. Iraji and B. Akhbari for comments that improved the presentation. 
        
\end{document}